\documentstyle[cite]{article}

\newcommand{\Sv}{{\bf S}}
\newcommand{\Rv}{{\bf R}}

\newcommand{\action}{\mbox{$\cal A$}}
\newcommand{\actions}{\mbox{$\cal A_{\mbox{\tiny FP}}$}}
\newcommand{\trkern}{\mbox{$\cal T$}} 
\newcommand{\Bsite}{{\mbox{\tiny B}}}

\newcommand{\half}{\mbox{$\scriptstyle \frac{1}{2}$}}

\catcode`@=11
\def\@sim#1#2{\setbox0=\hbox{$\sim$}\lower.9\ht0\vbox{\baselineskip0pt
              \lineskip0.1ex\ialign{$\m@th#1\hfill##\hfill$\crcr#2\crcr
              \sim\crcr}}}
\def\lsim{\mathrel{\mathpalette\@sim<}}
\def\gsim{\mathrel{\mathpalette\@sim>}}
\catcode`@=12

\newcommand{\reell}
    {{\kern+.25em\sf{R}\kern-.78em\sf{I}\kern+.78em\kern-.25em}}
\newcommand{\komplex}
    {{\sf{C}\kern-.46em\sf{I}\kern+.46em\kern-.25em}}
\newcommand{\posganz}
    {{\kern+.25em\sf{N}\kern-.86em\sf{I}\kern+.86em\kern-.25em}}
\newcommand{\ganz}
    {{\kern+.25em\sf{Z}\kern-.78em\sf{Z}\kern+.78em\kern-.65em}}
\newcommand{\Hamilton}
    {{\kern+.25em\sf{H}\kern-.86em\sf{I}\kern+.86em\kern-.25em}}
\newcommand{\Cayley}
    {{\sf{O}\kern-.56em\sf{I}\kern+.56em\kern-.25em}}
\newcommand{\unit}
    {{\sf{1}\kern-.18em\sf{I}\kern+.18em\kern-.18em}}
\newcommand{\opunit}
    {{\sf{1}\kern-.29em\sf{1}\kern+.29em\kern-.33em}}

\begin{document}

\input epsf

\hyphenation{in-vari-ant}

\thispagestyle{empty}
\parskip=12pt
\raggedbottom

\noindent
\hspace*{10cm} BUTP--95/17\\
\vspace*{1cm}
\begin{center}
{\LARGE Instantons and the fixed point topological charge in the
two--dimensional O(3) $\sigma$--model}
\footnote{Work supported in part by Schweizerischer Nationalfonds}

\vspace{1cm}

Marc Blatter,
Rudolf Burkhalter,
Peter Hasenfratz and
Ferenc Niedermayer\footnote{On leave from the Institute of Theoretical
Physics, E\"{o}tv\"{o}s University, Budapest}
\\
Institut f\"{u}r theoretische Physik \\
Universit\"{a}t Bern \\
Sidlerstrasse 5, CH--3012 Bern, Switzerland

\vspace{1.5cm}

{August 1995} \vspace*{0.5cm}          

\begin{abstract}
We define a fixed point topological charge for the two--dimensional
O(3) lattice $\sigma$--model which is free of topological defects.  We
use this operator in combination with the fixed point action to 
measure the topological susceptibility for a wide range of correlation
lengths.  The results strongly suggest that it is not a physical
quantity in this model. The procedure, however, can be applied to
other asymptotically free theories as well.

\end{abstract}

\end{center}

\vspace*{3cm}

\newpage 

\section{Introduction}
\label{intro}

Topological effects play an important role in the dynamics of
asymptotically free field theories.  In QCD instantons may be
responsible for breaking the axial symmetry resolving the so--called
U(1) problem \cite{WEINBERG}.  In a large $N_c$ limit the
topological susceptibility relates the masses of the pseudo--scalars
$\eta$, $\eta'$ and $K$ \cite{WITTEN}. 

The topological susceptibility $\chi_t$ may be defined as the infinite
volume limit of
\begin{equation}
\chi^{V}_t = \frac{\langle Q^2\rangle}{V},
\end{equation}
where $Q$ is the topological charge and $V$ is the space--time volume.
In the two--dimensional O(3) non--linear $\sigma$--model it is a
dimension two quantity that vanishes to all orders in the weak
coupling expansion. From the perturbative renormalization group (RG)
it is expected to scale according to the two loop $\beta$ function
\begin{equation} \label{eq-scaling}
\chi_t \propto \beta^2 \exp(- 4 \pi \beta), \qquad
( \beta \rightarrow \infty).
\end{equation}

It is a non--trivial task to recover the correct continuum results from
lattice Monte Carlo simulations.  A lattice topological charge
definition is needed which returns even for large fluctuations
reliable results.

A `geometric' definition proposed by Berg and L\"uscher \cite{BERG1}
is based on adding up the area of spherical triangles which are
defined by the spin vectors in an elementary plaquette. As the
contributions from all plaquettes are summed up, the internal
space --- the sphere described by the spin variables --- is covered and
if periodic boundary conditions are used one obtains an integer charge
signifying the number of times this sphere is `wrapped'.  The
topological susceptibility evaluated with this charge definition (and
the standard action) completely failed to scale
\cite{BERG1,LUSCHER1,BERG2}.  The reason was ascribed to special
configurations called `dislocations' \cite{BERG2}, which are dominant
in the statistical average. Dislocations are non--zero charged
configurations whose contributions to the topological charge come
entirely from small localized regions where they become `singular'.
If the minimal action of dislocations is smaller than the continuum
value of a one--instanton configuration (i.e. $4\pi$) then
dislocations will dominate the path integral and spoil the scaling
behaviour, eq.~(\ref{eq-scaling}) \cite{LUSCHER1,BERG2}.

Another definition goes back to DiVecchia et al.\ \cite{DIVECCHIA} ---
for a recent discussion including the fixed point (FP) action see
ref.~\cite{DELIA}.  It is called field theoretical or plaquette
definition and uses a `naive' discretization of the continuum charge
operator.  This prescription does not yield integer values and to
obtain continuum results renormalization factors are needed.  For
large $\beta$ these factors can be determined perturbatively, but for
intermediate $\beta$ one has to use non--perturbative techniques
\cite{TEPER1,DIGIACOMO,TEPER2,MICHAEL}.  Results obtained with the
field theoretical charge indicate for the susceptibility a behaviour
consistent with scaling
\cite{DIGIACOMO,MICHAEL,FARCHIONI1}.

A serious problem in these approaches is the role of the lattice
artifacts, sensitive both to the form of the lattice action and the
choice of the topological charge. A recent work \cite{HASENFRATZ1}
suggests to use the FP action of a renormalization group transformation
to study topological effects. In particular, an important feature is that
the FP action has scale invariant instanton solutions (with an action
value exactly $4\pi$), and hence --- as will be discussed in this
paper --- one can define a topological charge with no lattice
defects. In ref.~\cite{HASENFRATZ1} and here the O(3) $\sigma$--model
is considered, but the methods apply to other asymptotically free
theories as well. The SU(3) gauge theory has been studied in
refs.~\cite{DEGRAND1,DEGRAND2}.  A subsequent paper by one of us
\cite{BURKHALTER}, will deal with the application of these ideas to
$\mbox{CP}^{\mbox{\rm \tiny N-1}}$ models and in particular to the
$\mbox{CP}^{3}$ model.  Some of our results were already
presented in ref.~\cite{BLATTER}.  

The paper is organized as follows: First we review some results
derived in ref.~\cite{HASENFRATZ1} and define the FP field operator.
This is followed by a closer look at instantons in the continuum, in a
finite periodic volume and finally on a lattice using the FP
action. We then define the FP topological charge and present some
numerical results on classical solutions.  In the last section we
analyze the topological susceptibility evaluated in a Monte Carlo
simulation. After a brief description of the methods used, we present
the results which are followed by a conclusion and an outlook.

\section{RG results at the classical level}

\subsection{Review of the RG transformation and its fixed point}
\label{review}

Let us briefly summarize some RG results which were developed in a
previous paper \cite{HASENFRATZ1}.  For a detailed discussion we refer
the reader to this paper.

We consider the O(3) non--linear $\sigma$--model in two--dimensional
Euclidean space defined on a square lattice. The partition function
reads as follows
\begin{equation}
Z = \int\! \!  \mbox{D} {\rm \bf S}  \;  
e^{ -\beta {\cal A}( \hbox{\scriptsize \bf S} ) }.
\end{equation}
Here $ \mbox{D} \Sv $ is the O(3) invariant measure
\begin{equation}
\mbox{D} \Sv  = \prod_n d^{3} \mbox{S}_n \; \delta( \Sv_n^2 - 1)
\end{equation}
and $\beta \action(\Sv) $ is a regularization of the continuum action
\begin{equation}
\beta {\cal A}_{\mbox{\tiny cont}}(\Sv) = 
{ \frac{\beta}{2} } \int \! d^2 \! x \, 
 \partial_{\mu}  \Sv(x) \, \partial_{\mu}  \Sv(x),
\qquad \mbox{where} \qquad \Sv^2(x) = 1.
\end{equation}

We perform exact RG transformations by a Kadanoff type of blocking,
i.e.\ we divide the lattice into $2 \times 2$ blocks labeled by
indices $n_\Bsite$.  To each block we define a block spin variable
$\Rv_{n_\Bsite}$ which is some mean of the spin variables $\Sv_n$ in
the block.  The block spins $\Rv_{n_\Bsite}$ form a lattice
whose spacing is twice as large as the original one. An effective
action is defined by integration over the original lattice:
\begin{equation}
 e^{-\beta' {\cal A}'( \mbox{\scriptsize \bf R} )} = \int\! \!
\mbox{D} \Sv \, e^{ -\beta \left [ {\cal A}(\mbox{\scriptsize \bf S})
+ {\cal T}(\mbox{\scriptsize \bf R},\mbox{\scriptsize \bf S} ) \right
] },
\end{equation}
where ${\cal T}$ is the kernel of the RG transformation and its
normalization 
\begin{equation}
\int\! \!  \mbox{D} \Rv  \, e^{ -\beta 
{\cal T}(\mbox{\scriptsize \bf R},\mbox{\scriptsize \bf S} ) } = 1
\end{equation}
ensures the invariance of the partition function under this
transformation.  In the classical limit $(\beta \rightarrow \infty)$
the path integral is dominated by its saddle point:
\begin{equation}
{\cal A}'(\Rv) = \min_{\{ \mbox{\scriptsize \bf S} \} } 
\left \{  {\cal A}(\Sv) + {\cal T}(\Rv,\Sv) \right\}.
\end{equation}
The transformation kernel used in ref.~\cite{HASENFRATZ1} has in the
limit $\beta \rightarrow \infty$ the simple form:
\begin{equation}
{\cal T}(\Rv,\Sv) = \kappa \sum_{n_\Bsite} \left ( 
\left | \sum_{n \in n_\Bsite}\Sv_n \right | - \Rv_{n_\Bsite} \cdot 
\sum_{n\in n_\Bsite}\Sv_n \right ).
\end{equation}
Here $\kappa$ is a free positive parameter of the RG transformation,
which is tuned to make the FP action as compact as possible. As
indicated by the free field theory in one dimension, the choice
$\kappa = 2$ gives the most short--ranged FP action. A fixed point of
the transformation satisfies the equation
\begin{equation}  
\label{FP-equation}
{\cal A}_{\mbox{\tiny FP}}(\Rv) = \min_{\{ \mbox{\scriptsize \bf S} \} } 
\left \{  {\cal A}_{\mbox{\tiny FP}}(\Sv) + {\cal T}(\Rv,\Sv) \right \}.
\end{equation}
This equation --- called FP equation --- fixes for arbitrary
configurations $\{\Rv\}$ the value of the FP action.  Starting from a
lattice regularization of the continuum action repeated RG
transformations will drive the effective action to its fixed point.  
This takes
on the form of a minimization in a multigrid of lattice
configurations: 
\begin{equation}  \label{iFP-equation}
\action^{(k)}(\Rv) = \min_{\{ \mbox{\scriptsize \bf
 S}^{(1)},\mbox{\scriptsize \bf S}^{(2)},\ldots, \mbox{\scriptsize \bf
 S}^{(k)} \}} \left \{ \action^{(0)}(\Sv^{(k)}) +
 \trkern(\Sv^{(k)},\Sv^{(k-1)}) + \ldots + \trkern(\Rv,\Sv^{(1)})
 \right \}.
\end{equation}

\begin{figure}[htb]
\begin{center}
\vskip -5mm
\leavevmode
\epsfxsize=90mm
\epsfbox{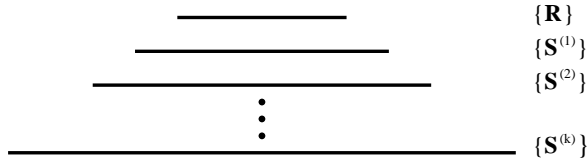}
\vskip 5mm
\caption{A multigrid is obtained by iterating the FP equation.}
\label{levels}
\end{center}
\end{figure}
On each successive level --- see figure~\ref{levels} --- the spin
configurations become smoother, not only because the lattice spacing
is halved, but also because the minimization tends to smooth out the
fluctuations around a solution to the equations of motion.  Hence one
may choose for the action $\action^{(0)}(\Sv^{(k)})$ on the finest
configuration $\{\Sv^{(k)}\}$ any lattice discretization of the
continuum action. The FP action $\actions$ is then obtained as the
limit of $k \rightarrow \infty$ of $\action^{(k)}(\Rv)$. For practical
purposes, however, only a few levels are needed and starting from the
standard action on the lowest level the FP value is reached soon.

\subsection{Parametrization of the FP action}
\label{paraction}

In principle the above multigrid approach can be used to evaluate the
FP action for arbitrary configurations to any precision desired.  For
practical calculations, however, a parametrization of the FP action is
needed.  In ref.~\cite{HASENFRATZ1} a parametrization has been
obtained by fitting the known values of the action for $\sim 500$
configurations. That parametrization represented well the FP action on
those configurations.  However, to control the topological effects
better, we decided to improve the parametrization further by
including some small size topological solutions in the fitting
procedure. We used several two--instanton solutions of the lattice FP
action.  While improving the fit for these instanton solutions, the
new parametrization does not affect the quality of the fit for the
previous configurations.

The resulting couplings are given in table~\ref{newcouplings} --- together
with a graphical notation of the corresponding operators. Let us explain
here again the meaning of this notation. The
parametrization of the action has the form
\begin{equation}
{\cal A}_{\mbox{\tiny FP}}(\Sv) = \sum \; {\rm coupling \; \times \;
products \; of} \; \frac{1}{2}\vartheta_{n_i,n_j}^2\,,
\end{equation}
where $\vartheta_{n_i,n_j}$ is the angle between the two spins
$\Sv_{n_i}$ and $\Sv_{n_j}$. Two dots connected with a line
\begin{picture}(22,6)(0,0) 
\put(2,3){\circle*{3.5}} 
\put(20,3){\circle*{3.5}} 
\put(2,3){\line(1,0){18}}
\end{picture} 
represent a factor $\frac{1}{2}\vartheta_{n_i,n_j}^2$ in the
action and the positions of the dots represent the lattice sites $n_i$
and $n_j$ respectively. Double, triple connected dots stand for the
square, cube of the above factors. The operator, finally, is the
product of all the factors $\frac{1}{2}\vartheta_{n_i,n_j}^2$
as indicated by the lines in the figure. The quadratic
and quartic couplings \# 1,2,4,5,7,10,16 and 19 are determined
analytically \cite{HASENFRATZ1}, the others were determined with a
numerical fitting procedure with the new instanton configurations
added. 
\begin{table} 
\setlength{\unitlength}{0.5mm}
\begin{tabular}{r cr c cr c cr}
\# & type & coupling & ~~~ & type & coupling
& ~~~ & type & coupling \\

1 &                         
\begin{picture}(20,10)(0,3)
\put(5,5){\circle*{2}}
\put(15,5){\circle*{2}}
\put(5,5){\line(1,0){10}}
\end{picture} & $0.61884$ & &

\begin{picture}(20,10)(0,3)
\put(5,5){\circle*{2}}
\put(15,5){\circle*{2}}
\put(5,4.5){\line(1,0){10}}
\put(5,5.5){\line(1,0){10}}
\end{picture} & $-0.04957$ & &

\begin{picture}(20,10)(0,3)
\put(5,5){\circle*{2}}
\put(15,5){\circle*{2}}
\put(5,4.1){\line(1,0){10}}
\put(5,5){\line(1,0){10}}
\put(5,5.9){\line(1,0){10}}
\end{picture} & $-0.00932$ \\

4 &                        
\begin{picture}(20,14)(0,5)
\put(5,1){\circle*{2}}
\put(15,11){\circle*{2}}
\put(5,1){\line(1,1){10}}
\end{picture} & $0.19058$  & &

\begin{picture}(20,14)(0,5)
\put(5,1){\circle*{2}}
\put(15,11){\circle*{2}}
\put(5,0.5){\line(1,1){10}}
\put(5,1.5){\line(1,1){10}}
\end{picture} & $-0.02212$  & &

\begin{picture}(20,14)(0,5)
\put(5,1){\circle*{2}}
\put(15,11){\circle*{2}}
\put(5,0.1){\line(1,1){10}}
\put(5,1.9){\line(1,1){10}}
\put(5,1){\line(1,1){10}}
\end{picture} & $-0.00746$ \\

7 &                        
\begin{picture}(20,14)(0,5)
\put(5,1){\circle*{2}}
\put(15,1){\circle*{2}}
\put(15,11){\circle*{2}}
\put(5,1){\line(1,0){10}}
\put(5,1){\line(1,1){10}}
\end{picture} & $0.01881$  & &

\begin{picture}(20,14)(0,5)
\put(5,1){\circle*{2}}
\put(15,1){\circle*{2}}
\put(15,11){\circle*{2}}
\put(5,0.5){\line(1,0){10}}
\put(5,1.5){\line(1,0){10}}
\put(5,1){\line(1,1){10}}
\end{picture} & $-0.00180$  & &

\begin{picture}(20,14)(0,5)
\put(5,1){\circle*{2}}
\put(15,1){\circle*{2}}
\put(15,11){\circle*{2}}
\put(5,1){\line(1,0){10}}
\put(5,0.5){\line(1,1){10}}
\put(5,1.5){\line(1,1){10}}
\end{picture} & $0.00658$ \\

10 &                          
\begin{picture}(20,14)(0,5)
\put(5,1){\circle*{2}}
\put(15,1){\circle*{2}}
\put(15,11){\circle*{2}}
\put(5,1){\line(1,0){10}}
\put(15,1){\line(0,1){10}}
\end{picture} & $0.02155$  & &  

\begin{picture}(20,14)(0,5)
\put(5,1){\circle*{2}}
\put(15,1){\circle*{2}}
\put(15,11){\circle*{2}}
\put(5,0.5){\line(1,0){10}}
\put(5,1.5){\line(1,0){10}}
\put(15,1){\line(0,1){10}}
\end{picture} & $0.00536$  & &  

\begin{picture}(20,14)(0,5)
\put(5,1){\circle*{2}}
\put(15,1){\circle*{2}}
\put(15,11){\circle*{2}}
\put(5,0.5){\line(1,0){10}}
\put(5,1.5){\line(1,0){10}}
\put(14.5,1){\line(0,1){10}}
\put(15.5,1){\line(0,1){10}}
\end{picture} & $-0.00081$ \\

13 &                        
\begin{picture}(20,14)(0,5)
\put(5,1){\circle*{2}}
\put(15,1){\circle*{2}}
\put(15,11){\circle*{2}}
\put(5,1){\line(1,0){10}}
\put(5,1){\line(1,1){10}}
\put(15,1){\line(0,1){10}}
\end{picture} & $0.00941$  & &

\begin{picture}(20,14)(0,5)
\put(5,1){\circle*{2}}
\put(15,1){\circle*{2}}
\put(15,11){\circle*{2}}
\put(5,0.5){\line(1,0){10}}
\put(5,1.5){\line(1,0){10}}
\put(5,1){\line(1,1){10}}
\put(15,1){\line(0,1){10}}
\end{picture} & $0.00488$  & &

\begin{picture}(20,14)(0,5)
\put(5,1){\circle*{2}}
\put(15,1){\circle*{2}}
\put(15,11){\circle*{2}}
\put(5,1){\line(1,0){10}}
\put(5,0.5){\line(1,1){10}}
\put(5,1.5){\line(1,1){10}}
\put(15,1){\line(0,1){10}}
\end{picture} & $-0.00225$ \\

16 &                        
\begin{picture}(20,14)(0,5)
\put(5,1){\circle*{2}}
\put(15,1){\circle*{2}}
\put(15,11){\circle*{2}}
\put(5,11){\circle*{2}}
\put(5,1){\line(1,1){10}}
\put(5,11){\line(1,-1){10}}
\end{picture} & $0.01209$  & &

\begin{picture}(20,14)(0,5)
\put(5,1){\circle*{2}}
\put(15,1){\circle*{2}}
\put(15,11){\circle*{2}}
\put(5,11){\circle*{2}}
\put(5,0.5){\line(1,1){10}}
\put(5,1.5){\line(1,1){10}}
\put(5,11){\line(1,-1){10}}
\end{picture} & $0.00534$  & &

\begin{picture}(20,14)(0,5)
\put(5,1){\circle*{2}}
\put(15,1){\circle*{2}}
\put(15,11){\circle*{2}}
\put(5,11){\circle*{2}}
\put(5,0.5){\line(1,1){10}}
\put(5,1.5){\line(1,1){10}}
\put(5,10.5){\line(1,-1){10}}
\put(5,11.5){\line(1,-1){10}}
\end{picture} & $0.00066$ \\

19 &                      
\begin{picture}(20,14)(0,5)
\put(5,1){\circle*{2}}
\put(15,1){\circle*{2}}
\put(15,11){\circle*{2}}
\put(5,11){\circle*{2}}
\put(5,1){\line(1,0){10}}
\put(5,11){\line(1,0){10}}
\end{picture} & $-0.00258$  & &

\begin{picture}(20,14)(0,5)
\put(5,1){\circle*{2}}
\put(15,1){\circle*{2}}
\put(15,11){\circle*{2}}
\put(5,11){\circle*{2}}
\put(5,0.5){\line(1,0){10}}
\put(5,1.5){\line(1,0){10}}
\put(5,11){\line(1,0){10}}
\end{picture} & $-0.00173$  & &

\begin{picture}(20,14)(0,5)
\put(5,1){\circle*{2}}
\put(15,1){\circle*{2}}
\put(15,11){\circle*{2}}
\put(5,11){\circle*{2}}
\put(5,0.5){\line(1,0){10}}
\put(5,1.5){\line(1,0){10}}
\put(5,10.5){\line(1,0){10}}
\put(5,11.5){\line(1,0){10}}
\end{picture} & $0.00146$ \\

22 &                     
\begin{picture}(20,14)(0,5)
\put(5,1){\circle*{2}}
\put(15,1){\circle*{2}}
\put(15,11){\circle*{2}}
\put(5,11){\circle*{2}}
\put(5,1){\line(1,0){10}}
\put(5,11){\line(1,0){10}}
\put(5,1){\line(0,1){10}}
\end{picture} & $-0.01040$  & &

\begin{picture}(20,14)(0,5)
\put(5,1){\circle*{2}}
\put(15,1){\circle*{2}}
\put(15,11){\circle*{2}}
\put(5,11){\circle*{2}}
\put(5,1){\line(1,0){10}}
\put(5,11){\line(1,0){10}}
\put(4.5,1){\line(0,1){10}}
\put(5.5,1){\line(0,1){10}}
\end{picture} & $-0.00218$  & &

\begin{picture}(20,14)(0,5)
\put(5,1){\circle*{2}}
\put(15,1){\circle*{2}}
\put(15,11){\circle*{2}}
\put(5,11){\circle*{2}}
\put(5,1){\line(1,0){10}}
\put(5,11){\line(1,0){10}}
\put(5,1){\line(0,1){10}}
\put(15,1){\line(0,1){10}}
\end{picture} & $0.02720$  \\

\end{tabular}
\caption{Couplings used for the parametrization of the FP action
including instanton configurations. The graphical notation is
explained in the text.}
\label{newcouplings}
\end{table}

\subsection{The fixed point field}
\label{parfield}

As we shall see in section~\ref{tcharge}, 
FP operators are closely related to the FP field.
The FP field is the fine field $\Sv^{(k)}$ in the
multigrid solution of the iterated FP equation~(\ref{iFP-equation}) as
$k$ goes to infinity.  If the functional dependence of the solution on
the first fine level $\Sv^{(1)}$ on $\Rv$ is known, the FP
field can be evaluated by iteration.  Below we construct the operator
$\Sv^{(1)}=\Sv^{(1)}(\Rv)$. (In the limit $k \rightarrow \infty$ the
solution $\{ \Sv^{(1)} \}$ of the iterated FP equation is identical to
the solution $\{ \Sv \}$ of the FP equation.)

For smooth fields a quadratic approximation of the FP equation 
can be made which can be solved analytically. 
Consider a smooth configuration $\{\Rv\}$
where the spins fluctuate around the first axis:
\begin{equation} \label{13}
\Rv_{n_\Bsite}=\left( {\sqrt{1-\vec{\chi}_{n_\Bsite}^{\,2}}
\atop \vec{\chi}_{n_\Bsite}} \right),
\end{equation}
where $\vec{\chi}_{n_\Bsite}$ has $2$ components, and
$|\vec{\chi}_{n_\Bsite}|\ll 1$.
For the minimizing fine field we can make the ansatz
\begin{equation} \label{14}
\Sv_{n}^{(1)}=\left( {\sqrt{1-\vec{\pi}_{n}^{\,2}}
\atop \vec{\pi}_{n}} \right).
\end{equation}
Inserted in the FP equation, the above expansion leads in leading
order to the free field case. This was solved by Bell and Wilson
\cite{BELL} --- for a brief review see for instance
ref.~\cite{HASENFRATZ1} --- here we only report the relation between
$\vec{\pi}_n$ and $\vec{\chi}_{n_{\mbox{\tiny B}}}$:
\begin{equation} \label{15}
\vec{\pi}_n = \sum_{n_\Bsite} \alpha (n, n_\Bsite)
\vec{\chi}_{n_{\mbox{\tiny B}}}.
\end{equation}
Here $\alpha$ is given by
\begin{equation} 
\alpha (n, n_{\mbox{\tiny B}}) = 
\int_0^{2\pi} \frac{d^2q}{(2\pi)^2}
e^{- i q ( n - 2 n_{\mbox{\tiny B}} ) } 
\frac{1}{4} \frac{\tilde{\rho}_{\mbox{\tiny FP}}(2 q)}
                 {\tilde{\rho}_{\mbox{\tiny FP}}( q)}
 \prod_{j = 1}^2 \frac{1 - e^{- 2 i q_j } }
                      {1 - e^{-  i q_j } }
\end{equation}
where $\tilde{\rho}_{\mbox{\tiny FP}} (q)$ is the coefficient in the
quadratic part of the FP action (given by the free field case).  By
iterating eq.~(\ref{15}) one can obtain the FP field operator in the free
field case.  The form is very similar to the above result, but with
slightly modified parameters \cite{DEGRAND1}.

On smooth configurations  $\{\Rv\}$ eqs.~(\ref{13}--\ref{15}) give
a good approximation. However we want to evaluate the fine field 
using a parametrization that performs well not only for smooth 
configurations $\{\Rv\}$, but also for general ones.
Our experience with the parametrization of the FP action suggests
the ansatz
\begin{equation}
\Sv_n = {\cal N} \left [\sum_{n_\Bsite} \alpha(n, n_\Bsite)
     \Rv_{n_\Bsite} + \!\!\!  \sum_{n_{\mbox{\tiny B}} \atop
     m_\Bsite^{},m_{\Bsite}'} \!\!\!  \beta(n,n_{\mbox{\tiny
     B}},m_{\Bsite}^{},m_{\mbox{\tiny B}}') \frac{1}{2}
     \vartheta^2_{m_\Bsite^{},m_\Bsite'} \Rv_{n_\Bsite} \right ],
\label{parametfield}
\end{equation}
where $\cal N$ is a normalizing factor which ensures $\Sv_n^2=1$. Like
for the action it proved to be useful to use instead of the scalar
product $(1 - \Rv_{m_\Bsite^{}}\Rv_{m_\Bsite'})$   
between the coarse spins at sites $m_\Bsite$ and $m_\Bsite'$ respectively,
the angle $\frac{1}{2}\vartheta^2_{m_\Bsite^{},m_\Bsite'}$.   
In order to
determine the coefficients $\beta$, we numerically minimized the FP
equation~(\ref{FP-equation}) using 60 configurations with lattice size
$5$ as input and stored the resulting fine lattices. The coefficients
were then determined by minimizing the difference between the
minimized fine spins and the parametrization (\ref{parametfield}).

The numerical values of the coefficients $\alpha$ and $\beta$ are
given in table~\ref{alphabeta} together with a symbolic notation of the
corresponding operators. We chose a set of 23 operators mainly because
of their compactness. \#~1--6 are the analytically determined
coefficients $\alpha$ and \#~7--23 are the numerically determined
coefficients $\beta$.  The meaning of the graphical notation of
the operators is the following: The dashed lines represent a $3\times
3$ section of the coarse lattice grid. The cross
\begin{picture}(8,6)(0,0)  
\put(4,3){\line(1,0){3.5}}
\put(4,3){\line(0,1){3.5}}
\put(4,3){\line(-1,0){3.5}}
\put(4,3){\line(0,-1){3.5}}
\end{picture} 
inbetween indicates the position $n$ of the fine spin $\Sv_n$ in
equation~(\ref{parametfield}). The little square
\begin{picture}(8,6)(0,0)  
\put(2,1){\framebox(4,4)} 
\end{picture} 
denotes the position $n_{\mbox{\tiny B}}$ of the coarse spin
$\Rv_{n_{\mbox{\tiny B}}}\,$. The two connected dots
\begin{picture}(22,6)(0,0) 
\put(2,3){\circle*{2.5}} 
\put(20,3){\circle*{2.5}} 
\put(2,3){\line(1,0){18}}
\end{picture} 
are the positions $m_{\mbox{\tiny B}}$ and $m_{\mbox{\tiny B}}'$ of
the spins whose angle $\frac{1}{2} \vartheta^2_{ m_{\mbox{\tiny
B}}^{}\, m_{\mbox{\tiny B}}'}$ enters into the parametrization.
Graphs obtained by trivial symmetry transformations are not drawn
separately.

\begin{table}\centering 
\setlength{\unitlength}{0.5mm}
\begin{tabular}{|c c c| c c c| c c c|}
\hline
\# & type & coeff & \# & type & coeff
& \# & type & coeff \\
\hline
1 &
\begin{picture}(20,20)(0,4)
\put(0,0){\dashbox{1}(10,10)}
\put(0,10){\dashbox{1}(10,10)}
\put(10,0){\dashbox{1}(10,10)}
\put(10,10){\dashbox{1}(10,10)}
\put(7.5,7.5){\line(1,0){1.5}}
\put(7.5,7.5){\line(0,1){1.5}}
\put(7.5,7.5){\line(-1,0){1.5}}
\put(7.5,7.5){\line(0,-1){1.5}}
\put(9,9){\framebox(2,2)}
\end{picture} & 
$0.59497$
&

2 &
\begin{picture}(20,20)(0,4)
\put(0,0){\dashbox{1}(10,10)}
\put(0,10){\dashbox{1}(10,10)}
\put(10,0){\dashbox{1}(10,10)}
\put(10,10){\dashbox{1}(10,10)}
\put(7.5,7.5){\line(1,0){1.5}}
\put(7.5,7.5){\line(0,1){1.5}}
\put(7.5,7.5){\line(-1,0){1.5}}
\put(7.5,7.5){\line(0,-1){1.5}}
\put(-1,9){\framebox(2,2)}
\end{picture} & 
$0.15621$
&

3 &
\begin{picture}(20,20)(0,4)
\put(0,0){\dashbox{1}(10,10)}
\put(0,10){\dashbox{1}(10,10)}
\put(10,0){\dashbox{1}(10,10)}
\put(10,10){\dashbox{1}(10,10)}
\put(7.5,7.5){\line(1,0){1.5}}
\put(7.5,7.5){\line(0,1){1.5}}
\put(7.5,7.5){\line(-1,0){1.5}}
\put(7.5,7.5){\line(0,-1){1.5}}
\put(-1,-1){\framebox(2,2)}
\end{picture} &  
$0.08300$ \\
 & & & & & & & & \\
\hline

4 &
\begin{picture}(20,20)(0,4)
\put(0,0){\dashbox{1}(10,10)}
\put(0,10){\dashbox{1}(10,10)}
\put(10,0){\dashbox{1}(10,10)}
\put(10,10){\dashbox{1}(10,10)}
\put(7.5,7.5){\line(1,0){1.5}}
\put(7.5,7.5){\line(0,1){1.5}}
\put(7.5,7.5){\line(-1,0){1.5}}
\put(7.5,7.5){\line(0,-1){1.5}}
\put(-1,19){\framebox(2,2)}
\end{picture} & 
$0.00942$
&

5 &
\begin{picture}(20,20)(0,4)
\put(0,0){\dashbox{1}(10,10)}
\put(0,10){\dashbox{1}(10,10)}
\put(10,0){\dashbox{1}(10,10)}
\put(10,10){\dashbox{1}(10,10)}
\put(7.5,7.5){\line(1,0){1.5}}
\put(7.5,7.5){\line(0,1){1.5}}
\put(7.5,7.5){\line(-1,0){1.5}}
\put(7.5,7.5){\line(0,-1){1.5}}
\put(9,19){\framebox(2,2)}
\end{picture} & 
$-0.00171$
&

6 &
\begin{picture}(20,20)(0,4)
\put(0,0){\dashbox{1}(10,10)}
\put(0,10){\dashbox{1}(10,10)}
\put(10,0){\dashbox{1}(10,10)}
\put(10,10){\dashbox{1}(10,10)}
\put(7.5,7.5){\line(1,0){1.5}}
\put(7.5,7.5){\line(0,1){1.5}}
\put(7.5,7.5){\line(-1,0){1.5}}
\put(7.5,7.5){\line(0,-1){1.5}}
\put(19,19){\framebox(2,2)}
\end{picture} &  
$-0.00668$ \\
 & & & & & & & & \\
\hline

7 &
\begin{picture}(20,20)(0,4)
\put(0,0){\dashbox{1}(10,10)}
\put(0,10){\dashbox{1}(10,10)}
\put(10,0){\dashbox{1}(10,10)}
\put(10,10){\dashbox{1}(10,10)}
\put(7.5,7.5){\line(1,0){1.5}}
\put(7.5,7.5){\line(0,1){1.5}}
\put(7.5,7.5){\line(-1,0){1.5}}
\put(7.5,7.5){\line(0,-1){1.5}}
\put(9,9){\framebox(2,2)}
\put(10,10){\circle*{1.5}}
\put(0,10){\circle*{1.5}}
\put(0,10){\line(1,0){10}}
\end{picture} & 
$-0.01228$
&

8 &
\begin{picture}(20,20)(0,4)
\put(0,0){\dashbox{1}(10,10)}
\put(0,10){\dashbox{1}(10,10)}
\put(10,0){\dashbox{1}(10,10)}
\put(10,10){\dashbox{1}(10,10)}
\put(7.5,7.5){\line(1,0){1.5}}
\put(7.5,7.5){\line(0,1){1.5}}
\put(7.5,7.5){\line(-1,0){1.5}}
\put(7.5,7.5){\line(0,-1){1.5}}
\put(9,9){\framebox(2,2)}
\put(10,10){\circle*{1.5}}
\put(0,0){\circle*{1.5}}
\put(0,0){\line(1,1){10}}
\end{picture} & 
$-0.02004$
&

9 &
\begin{picture}(20,20)(0,4)
\put(0,0){\dashbox{1}(10,10)}
\put(0,10){\dashbox{1}(10,10)}
\put(10,0){\dashbox{1}(10,10)}
\put(10,10){\dashbox{1}(10,10)}
\put(7.5,7.5){\line(1,0){1.5}}
\put(7.5,7.5){\line(0,1){1.5}}
\put(7.5,7.5){\line(-1,0){1.5}}
\put(7.5,7.5){\line(0,-1){1.5}}
\put(9,9){\framebox(2,2)}
\put(10,0){\circle*{1.5}}
\put(0,10){\circle*{1.5}}
\put(0,10){\line(1,-1){10}}
\end{picture} & 
$-0.03832$ \\
 & & & & & & & & \\
\hline

10 &
\begin{picture}(20,20)(0,4)
\put(0,0){\dashbox{1}(10,10)}
\put(0,10){\dashbox{1}(10,10)}
\put(10,0){\dashbox{1}(10,10)}
\put(10,10){\dashbox{1}(10,10)}
\put(7.5,7.5){\line(1,0){1.5}}
\put(7.5,7.5){\line(0,1){1.5}}
\put(7.5,7.5){\line(-1,0){1.5}}
\put(7.5,7.5){\line(0,-1){1.5}}
\put(9,9){\framebox(2,2)}
\put(0,10){\circle*{1.5}}
\put(0,0){\circle*{1.5}}
\put(0,0){\line(0,1){10}}
\end{picture} & 
$-0.05095$
&

11 &
\begin{picture}(20,20)(0,4)
\put(0,0){\dashbox{1}(10,10)}
\put(0,10){\dashbox{1}(10,10)}
\put(10,0){\dashbox{1}(10,10)}
\put(10,10){\dashbox{1}(10,10)}
\put(7.5,7.5){\line(1,0){1.5}}
\put(7.5,7.5){\line(0,1){1.5}}
\put(7.5,7.5){\line(-1,0){1.5}}
\put(7.5,7.5){\line(0,-1){1.5}}
\put(-1,9){\framebox(2,2)}
\put(10,10){\circle*{1.5}}
\put(0,10){\circle*{1.5}}
\put(0,10){\line(1,0){10}}
\end{picture} & 
$0.01475$
&

12 &
\begin{picture}(20,20)(0,4)
\put(0,0){\dashbox{1}(10,10)}
\put(0,10){\dashbox{1}(10,10)}
\put(10,0){\dashbox{1}(10,10)}
\put(10,10){\dashbox{1}(10,10)}
\put(7.5,7.5){\line(1,0){1.5}}
\put(7.5,7.5){\line(0,1){1.5}}
\put(7.5,7.5){\line(-1,0){1.5}}
\put(7.5,7.5){\line(0,-1){1.5}}
\put(-1,9){\framebox(2,2)}
\put(10,10){\circle*{1.5}}
\put(10,0){\circle*{1.5}}
\put(10,0){\line(0,1){10}}
\end{picture} & 
$-0.00586$ \\
 & & & & & & & & \\
\hline

13 &
\begin{picture}(20,20)(0,4)
\put(0,0){\dashbox{1}(10,10)}
\put(0,10){\dashbox{1}(10,10)}
\put(10,0){\dashbox{1}(10,10)}
\put(10,10){\dashbox{1}(10,10)}
\put(7.5,7.5){\line(1,0){1.5}}
\put(7.5,7.5){\line(0,1){1.5}}
\put(7.5,7.5){\line(-1,0){1.5}}
\put(7.5,7.5){\line(0,-1){1.5}}
\put(-1,9){\framebox(2,2)}
\put(10,10){\circle*{1.5}}
\put(0,0){\circle*{1.5}}
\put(0,0){\line(1,1){10}}
\end{picture} & 
$-0.00303$
&

14 &
\begin{picture}(20,20)(0,4)
\put(0,0){\dashbox{1}(10,10)}
\put(0,10){\dashbox{1}(10,10)}
\put(10,0){\dashbox{1}(10,10)}
\put(10,10){\dashbox{1}(10,10)}
\put(7.5,7.5){\line(1,0){1.5}}
\put(7.5,7.5){\line(0,1){1.5}}
\put(7.5,7.5){\line(-1,0){1.5}}
\put(7.5,7.5){\line(0,-1){1.5}}
\put(-1,9){\framebox(2,2)}
\put(10,0){\circle*{1.5}}
\put(0,10){\circle*{1.5}}
\put(0,10){\line(1,-1){10}}
\end{picture} & 
$-0.00123$
&

15 &
\begin{picture}(20,20)(0,4)
\put(0,0){\dashbox{1}(10,10)}
\put(0,10){\dashbox{1}(10,10)}
\put(10,0){\dashbox{1}(10,10)}
\put(10,10){\dashbox{1}(10,10)}
\put(7.5,7.5){\line(1,0){1.5}}
\put(7.5,7.5){\line(0,1){1.5}}
\put(7.5,7.5){\line(-1,0){1.5}}
\put(7.5,7.5){\line(0,-1){1.5}}
\put(-1,9){\framebox(2,2)}
\put(0,10){\circle*{1.5}}
\put(0,0){\circle*{1.5}}
\put(0,0){\line(0,1){10}}
\end{picture} & 
$-0.00118$ \\
 & & & & & & & & \\
\hline

16 &
\begin{picture}(20,20)(0,4)
\put(0,0){\dashbox{1}(10,10)}
\put(0,10){\dashbox{1}(10,10)}
\put(10,0){\dashbox{1}(10,10)}
\put(10,10){\dashbox{1}(10,10)}
\put(7.5,7.5){\line(1,0){1.5}}
\put(7.5,7.5){\line(0,1){1.5}}
\put(7.5,7.5){\line(-1,0){1.5}}
\put(7.5,7.5){\line(0,-1){1.5}}
\put(-1,9){\framebox(2,2)}
\put(10,0){\circle*{1.5}}
\put(0,0){\circle*{1.5}}
\put(0,0){\line(1,0){10}}
\end{picture} & 
$-0.01596$
&

17 &
\begin{picture}(20,20)(0,4)
\put(0,0){\dashbox{1}(10,10)}
\put(0,10){\dashbox{1}(10,10)}
\put(10,0){\dashbox{1}(10,10)}
\put(10,10){\dashbox{1}(10,10)}
\put(7.5,7.5){\line(1,0){1.5}}
\put(7.5,7.5){\line(0,1){1.5}}
\put(7.5,7.5){\line(-1,0){1.5}}
\put(7.5,7.5){\line(0,-1){1.5}}
\put(-1,-1){\framebox(2,2)}
\put(0,10){\circle*{1.5}}
\put(10,10){\circle*{1.5}}
\put(0,10){\line(1,0){10}}
\end{picture} & 
$-0.00090$
&

18 &
\begin{picture}(20,20)(0,4)
\put(0,0){\dashbox{1}(10,10)}
\put(0,10){\dashbox{1}(10,10)}
\put(10,0){\dashbox{1}(10,10)}
\put(10,10){\dashbox{1}(10,10)}
\put(7.5,7.5){\line(1,0){1.5}}
\put(7.5,7.5){\line(0,1){1.5}}
\put(7.5,7.5){\line(-1,0){1.5}}
\put(7.5,7.5){\line(0,-1){1.5}}
\put(-1,-1){\framebox(2,2)}
\put(10,10){\circle*{1.5}}
\put(0,0){\circle*{1.5}}
\put(0,0){\line(1,1){10}}
\end{picture} & 
$0.00140$ \\
 & & & & & & & & \\
\hline

19 &
\begin{picture}(20,20)(0,4)
\put(0,0){\dashbox{1}(10,10)}
\put(0,10){\dashbox{1}(10,10)}
\put(10,0){\dashbox{1}(10,10)}
\put(10,10){\dashbox{1}(10,10)}
\put(7.5,7.5){\line(1,0){1.5}}
\put(7.5,7.5){\line(0,1){1.5}}
\put(7.5,7.5){\line(-1,0){1.5}}
\put(7.5,7.5){\line(0,-1){1.5}}
\put(-1,-1){\framebox(2,2)}
\put(10,0){\circle*{1.5}}
\put(0,10){\circle*{1.5}}
\put(0,10){\line(1,-1){10}}
\end{picture} & 
$-0.00128$
&

20 &
\begin{picture}(20,20)(0,4)
\put(0,0){\dashbox{1}(10,10)}
\put(0,10){\dashbox{1}(10,10)}
\put(10,0){\dashbox{1}(10,10)}
\put(10,10){\dashbox{1}(10,10)}
\put(7.5,7.5){\line(1,0){1.5}}
\put(7.5,7.5){\line(0,1){1.5}}
\put(7.5,7.5){\line(-1,0){1.5}}
\put(7.5,7.5){\line(0,-1){1.5}}
\put(-1,-1){\framebox(2,2)}
\put(0,0){\circle*{1.5}}
\put(0,10){\circle*{1.5}}
\put(0,0){\line(0,1){10}}
\end{picture} & 
$0.00447$
&

21 &
\begin{picture}(20,20)(0,4)
\put(0,0){\dashbox{1}(10,10)}
\put(0,10){\dashbox{1}(10,10)}
\put(10,0){\dashbox{1}(10,10)}
\put(10,10){\dashbox{1}(10,10)}
\put(7.5,7.5){\line(1,0){1.5}}
\put(7.5,7.5){\line(0,1){1.5}}
\put(7.5,7.5){\line(-1,0){1.5}}
\put(7.5,7.5){\line(0,-1){1.5}}
\put(-1,19){\framebox(2,2)}
\put(0,10){\circle*{1.5}}
\put(10,10){\circle*{1.5}}
\put(0,10){\line(1,0){10}}
\end{picture} & 
$0.00199$ \\
 & & & & & & & & \\
\hline

22 &
\begin{picture}(20,20)(0,4)
\put(0,0){\dashbox{1}(10,10)}
\put(0,10){\dashbox{1}(10,10)}
\put(10,0){\dashbox{1}(10,10)}
\put(10,10){\dashbox{1}(10,10)}
\put(7.5,7.5){\line(1,0){1.5}}
\put(7.5,7.5){\line(0,1){1.5}}
\put(7.5,7.5){\line(-1,0){1.5}}
\put(7.5,7.5){\line(0,-1){1.5}}
\put(9,19){\framebox(2,2)}
\put(0,10){\circle*{1.5}}
\put(10,10){\circle*{1.5}}
\put(0,10){\line(1,0){10}}
\end{picture} & 
$0.00304$
&

23 &
\begin{picture}(20,20)(0,4)
\put(0,0){\dashbox{1}(10,10)}
\put(0,10){\dashbox{1}(10,10)}
\put(10,0){\dashbox{1}(10,10)}
\put(10,10){\dashbox{1}(10,10)}
\put(7.5,7.5){\line(1,0){1.5}}
\put(7.5,7.5){\line(0,1){1.5}}
\put(7.5,7.5){\line(-1,0){1.5}}
\put(7.5,7.5){\line(0,-1){1.5}}
\put(19,19){\framebox(2,2)}
\put(10,10){\circle*{1.5}}
\put(20,20){\circle*{1.5}}
\put(10,10){\line(1,1){10}}
\end{picture} & 
$-0.00015$
&

& & \\
 & & & & & & & & \\
\hline
\end{tabular}
\caption{Coefficients of the parametrization of the fine field.}
\label{alphabeta}
\end{table}

\section{Instantons}
\label{instanti}

\subsection{Instantons on the lattice}
\label{instanto}

In infinite volume continuum field theory topology is a well defined
concept.  Field configurations can be classified in topological
sectors according to a `winding number' or topological charge
\cite{BELAVIN2}. In a lattice formulation this concept breaks
down. When discretizing a theory, continuity in coordinate and
internal space is lost. On the other hand topology is based on
continuous transformations of mappings which are separable in
classes. In a discretized theory every field configuration can be
continuously transformed into any other.  If lattice configurations
are sufficiently smooth, an unambiguous topological charge may be
assigned.  Conversely, for field configurations containing large
fluctuations an interpolation is not unique and a charge definition
becomes ambiguous.

An additional problem arises due to the discretization of the
continuum action. While the continuum action possesses scale invariant
instanton solutions, this is generally not true for discretized
actions. In particular, starting with non--zero charged configurations
L\"uscher \cite{LUSCHER1} found that one can continuously lower the
standard lattice action to zero by a local minimization in the spin
variables.

In reference \cite{HASENFRATZ1} it was suggested to investigate the
above problems with the aid of renormalization group methods and it
was shown that the FP action has scale invariant instanton solutions.
In this section we continue along these lines and 
construct instanton solutions of the FP action.  These in turn can be
used to study the performance of a proposed improved topological
charge definition. But first let us review some basic facts about
instantons in the continuum \cite{BELAVIN2}.

In an infinite volume configurations with a finite action play a
special role: At `infinity' all spin variables point in the same
direction and the space $\reell^2$ can be compactified by stereographic
projection into a sphere $S^2$. A finite action configuration is thus
a mapping of a `coordinate' sphere onto the internal sphere
$\Sv_n^2=1$. Such mappings can be classified by homotopy classes, with
an integer number --- the topological charge $Q$ --- characterizing the
sectors.  While configurations from the same topological sector can be
continuously deformed into each other, this is not true for
configurations with a different charge.  The charge $Q$ is the number
of times the internal sphere is wrapped as the coordinate sphere is
traversed. It may be defined as the integral
\begin{equation}
Q= \frac{1}{8\pi}\int \! d^2 \! x \, \epsilon_{\mu\nu} \,
\Sv \cdot ( \partial_{\mu} \Sv \times \partial_{\nu} \Sv ).
\end{equation}
and it is related to the action by the inequality 
\begin{equation}
\action \geq 4\pi\, |Q|.
\label{actinequality}
\end{equation}
If for a given configuration the equality is satisfied, the
configuration minimizes the action and is therefore a solution of the
equations of motion.

Let us now turn to the theory in a finite volume. By demanding
periodic boundary conditions
\begin{equation}
\Sv(x_1 + Lm , x_2 + L n) = \Sv(x_1,x_2), \qquad \mbox{where} 
\qquad m, \, n \in \ganz,
\end{equation}
we define the theory on a square torus of size $L$.  In a finite
volume every field configuration has a finite action and due to the
periodic boundary conditions an integer topological charge $Q$
associated with it.

We can now explicitly construct pure instanton or pure anti--instanton
configurations with an action $\action=4\pi|Q|$ \cite{RICHARD}.  We
use the plane coordinates defined by the stereographic projection to
describe the solutions:
\begin{eqnarray}
\Sv_i & = & \frac{ 2 \, u_i}{ 1 + |u|^2} \hspace{1cm} i = 1,\, 2,
\nonumber \\ & & \\ \Sv_3 & = & \frac{1 - |u|^2}{1 + |u|^2}, \nonumber
\end{eqnarray}
where $u = u_1 + i u_2$ .
The instanton solutions at the boundary of equation~(\ref{actinequality}) 
satisfy the
Cauchy--Riemann equations for $u$ being an analytic function in 
$z = x_1 + i x_2$:
\begin{equation}
(\partial_1 + i \partial_2) u = 0.
\end{equation}

The solutions are doubly--periodic meromorphic functions called
elliptic functions \cite{RICHARD}.
They can be written as
\begin{equation}
u = c \prod_{i = 1}^{k} \frac{ \sigma(z-a_i) }{\sigma(z-b_i)}, 
\quad {\rm with} \quad \sum_{i=1}^{k} a_i = \sum_{i=1}^{k} b_i.
\end{equation}
Here the integer $k$ is the topological charge of the solution and
$c$, $a_1,\ldots,a_k$ and $b_1,\ldots,b_k$ are complex numbers.
$\sigma(z)$ is the Weierstrass $\sigma$--function with half--periods
$\omega = L / 2$ and $\omega'= iL / 2 $.  Using Cauchy's theorem one
can show that there are no solutions with topological charge equal to
one \cite{RICHARD}.  Hence we are forced to construct charge two
instanton solutions.  Specifically, we set $k=2$ and $c=1$.
A reasonable definition for the instanton size is 
\begin{equation}
\rho = \half \min\{|a_1-b_1|,|a_1-b_2|\}.
\label{size}
\end{equation}

Let us now turn to the construction of instanton solutions on the
lattice.  As was pointed out in ref.~\cite{HASENFRATZ1} the FP action
allows scale invariant instanton solutions.  Since we use this fact to
construct instantons on the lattice, it is appropriate to repeat the
statement:

If a given configuration $\{\Rv\}$ satisfies the equations of motion
for the FP action $\actions$ and it is a minimum, then the solution
$\{ \Sv \}$ of the FP equation~(\ref{FP-equation}) satisfies
the equations of motion also.  Moreover, both configurations yield the
same value for the action.

The proof is quite simple: If $\{\Rv\}$ is at a local minimum of the
FP action \actions, variations with respect to $\{\Rv\}$ will
vanish:
\begin{equation}
\frac{\delta\actions(\Rv)}{\delta\Rv_{n_\Bsite}} = 
\kappa \left ( - \sum_{n \in n_\Bsite} \Sv_n + \Rv_{n_\Bsite}
\left ( \Rv_{n_\Bsite} \cdot  \sum_{n \in n_\Bsite} \Sv_n \right  )
\right ) = 0.
\label{variation}
\end{equation}
Here $\Sv=\Sv(\Rv)$ is the solution of the FP equation with $\{ \Rv \}$
as coarse input configuration.
Since $\{ \Rv \}$ is a local minimum eq.~(\ref{variation}) implies
\begin{equation}
\Rv_{n_\Bsite} = \frac{\sum_{n \in n_\Bsite}\Sv_n}
                      {| \sum_{n \in n_\Bsite}\Sv_n | }.
\label{blocking}
\end{equation}
Consequently, we have
\begin{equation}
\trkern(\Rv,\Sv) = 0.
\end{equation}
Since the transformation kernel $\trkern(\Rv,\Sv) \ge 0$,
the configuration $\{ \Sv \}$  gives its minimum for fixed $\{ \Rv \}$.
Because the configuration $\{\Sv \}$ minimizes the right hand side 
of the FP equation, it minimizes the FP action $\actions(\Sv)$ separately.
Therefore it is a solution of the equations of motion. 
Since $\trkern(\Rv,\Sv) = 0$, both actions have the same value: 
$\actions(\Rv) = \actions(\Sv)$. This concludes the proof.

The reverse of the above statement is in general not true for arbitrary
configurations.  Although a fine configuration which is a solution of
the FP equations of motion and which is used to construct a coarse
configuration by means of the above blocking
equation~(\ref{blocking}), locally minimizes the right hand side of
the FP equation, this minimum does not necessarily coincide with the
absolute one. This in fact prevents the existence of arbitrary small
instanton solutions.

Using the above ideas it is clear how to construct instanton solutions of
\actions. We naively discretize the continuum instanton solution on a
very fine lattice with spacing $a_0=2^{-k}a$. After performing $k$
blocking steps as defined by equation~(\ref{blocking}), we obtain a
configuration $\{\Rv\}$ on a lattice with spacing $a$. In the limit $
k \rightarrow \infty $ we recover the continuum solution, which is of
course a local minimum of the continuum action. Since all the
successive blocked configurations minimize the transformation kernels
in the iterated FP equation, the configuration $\{ \Rv\}$ is a good
candidate for a lattice solution of the FP action\footnote{ It will be
a solution, unless the size of the instantons is too small with
respect to the lattice spacing.}. We may solve the FP equation for
$\{\Rv\}$ to check whether it is still a solution.  If $\{ \Rv \}$ is
a solution, then the multigrid minimization procedure should lead to
the same configurations on finer lattices as those which were used in
constructing $\{ \Rv \}$ by blocking.

\subsection{Definition of the topological charge  on the lattice}
\label{tcharge}

In the following we define a topological charge operator based on the
multigrid solution of the FP action.  We evaluate the FP topological
charge by means of the solutions of the FP
equation~(\ref{FP-equation}).  Under a RG transformation an operator
${ \cal O}( \Sv)$ transforms into ${\cal O}'(\Rv)$ on the coarse
lattice as
\begin{equation}
{\cal O}'(\Rv) e^{-\beta' {\cal A}'(\mbox{\scriptsize\bf R})} = \int\!
\! \mbox{D} \Sv \; 
{ \cal O}( \Sv) e^{ -\beta [ {\cal A}(\mbox{\scriptsize \bf
S}) + {\cal T}(\mbox{\scriptsize \bf R},\mbox{\scriptsize \bf S} ) ]
},
\end{equation}

In the limit $\beta \rightarrow \infty$ the path integral on the right
hand side is approximated by its saddle point and we obtain 
\begin{equation}
{\cal O}'(\Rv) = {\cal O}(\Sv(\Rv)),
\end{equation}
where the spin configuration $\{\Sv(\Rv)\}$ is the solution of the FP
equation~(\ref{FP-equation}).  Repeated application of this
transformation will single out the operator with largest eigenvalue.
Since the topological charge is expected to be a marginal operator, we
may obtain it as the limit 
\begin{equation}
Q_{\mbox{\rm \tiny FP}}(\Rv) = \lim_{k \rightarrow \infty}
Q(\Sv^{(k)}(\Rv)).
\end{equation}
Here $Q$ is some standard lattice charge definition and $\{ \Sv^{(k)}
\}$ is the solution of the iterated FP equation~(\ref{iFP-equation})
on the lowest level in a $k$ level multigrid (see figure~\ref{levels}).
In other words, the FP topological charge is a standard topological charge
evaluated on the FP field. 

Note that $\Sv^{(k)}(\Rv)$ becomes increasingly smooth as $k$ grows:
first, because the corresponding lattice spacing $a_0=2^{-k} a$
decreases, secondly, because $\Sv^{(k)}(\Rv)$ becomes almost a solution
to the equations of motion. Consequently, any sensible definition of the
topological charge can be used on these configurations, the final result
will not depend on this choice. Nevertheless, it is more convenient
to use the geometric definition since it is stable against small
variations of the field and hence $k$, the number of levels in the 
multigrid minimization could be kept small with this definition.
For a review of the geometric definition we refer the reader to 
ref.~\cite{BERG1}.

One can easily show that with this definition of the topological
charge there are no dangerous dislocations. More precisely,
one has
\begin{equation} \label{31}
{\cal A}_{\mbox{\rm \tiny FP}}(\Rv) \geq 4 \pi \, | Q_{\mbox{\rm \tiny
FP}}(\Rv) |,
\end{equation}
for arbitrary configuration $\{ \Rv \}$.
The corresponding statement is true in the continuum, hence it is also
true for $\Sv^{(k)}(\Rv)$ for $k\to\infty$. Eq.~(\ref{31}) follows 
then by observing that the contribution of the ${\cal T}$ terms in 
eq.~(\ref{FP-equation}) is non--negative.
We are now ready to discuss the numerical
aspects of classical solutions.

\subsection{Classical numerical results}
\label{classres}

\begin{figure}[htb]
\begin{center}
\leavevmode
\epsfxsize=110mm
\epsfbox{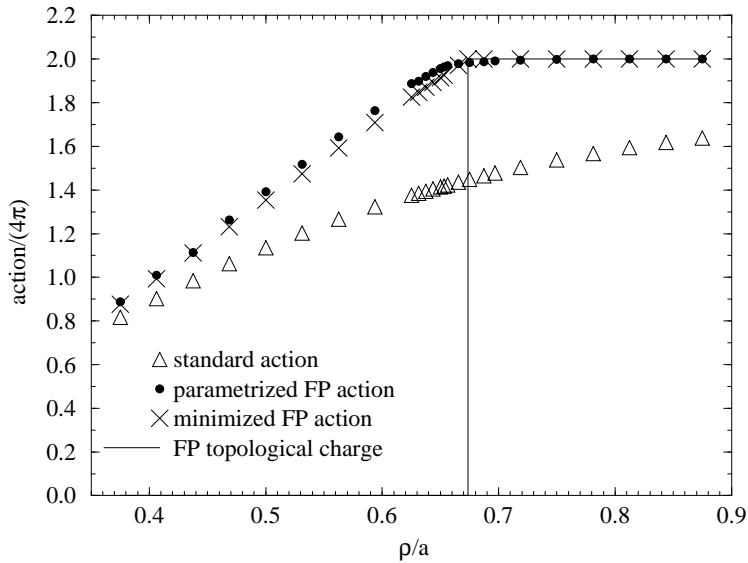}
\vskip -5mm
\caption{Actions and charge of instantons with radii of the order of one
lattice spacing}
\label{rhovar}
\end{center}
\end{figure}

Following the above program we naively discretize two--instanton
solutions on the torus of various sizes on very fine lattices.
We found that four to five blocking steps are sufficient to make any
lattice artifacts of the original discretization negligible.  
On the finally blocked configurations we can measure several
quantities. On the coarse configuration itself we measure the standard
action and the parametrization of the FP action presented in
section~\ref{paraction}. Performing a minimization on a multigrid with
three finer levels we measure the exact FP action and on the finest
level the FP charge.  Using the instanton radius given by
eq.~(\ref{size}), we get a parameter which characterizes well the
breakdown of the blocking to obtain instanton solutions on coarse
lattices: In figure~\ref{rhovar} it can be clearly seen that below an
instanton radius of $\rho \lsim 0.7 a$ the configurations are no
longer instanton solutions and we shall say that the instanton falls
through the lattice. It is gratifying to see that the FP charge
immediately falls off to zero, as the FP action drops below the
continuum value.  Furthermore, figure~\ref{rhovar} demonstrates how
well the parametrization for the FP action is suited for instanton
configurations. The deviation from the exact value is quite small, in
particular the parametrized FP action is only marginally smaller than
the continuum value in the region above the point where the instanton
falls through the lattice. In contrast, the values of the standard
action are quite different from the continuum ones. These numerical
findings support the above statement, that there are no dangerous
dislocations present when using the FP action together with the FP
topological charge.

In numerical simulations we use a parametrized form of the
FP action and the parametrization in eq.~(\ref{parametfield}) for the
FP field instead of the time consuming minimization procedure.  One
then is interested if the use of these two parametrizations has an
influence on the existence of dislocations.  The curves referring to
the FP action and FP charge in fig.~\ref{rhovar} have a non--analytic
break at $\rho / a \simeq 0.7$.  Our parametrization does not fully
reproduce this behaviour, and so one might expect that observable
deviations will occur in this instanton region.  We systematically
searched for minimal action configurations with a parametrized action
lower than the continuum value in the $Q_{\mbox{\tiny FP}}^{\rm
par.}=2$ sector.  We found a minimal action of $1.84\cdot 4\pi$
(compare this with the value $0.93\cdot 4\pi$ we found using the
standard action and geometric charge). This value can be ascribed to
the not exactly accurate parametrization of the FP field. If we
actually solve the FP equation (\ref{FP-equation}) for this
`dislocation configuration' we get the correct charge
$Q=0$. Nevertheless, as one has to use parametrizations for Monte
Carlo simulations, such configurations could be dangerous.  On the
other hand, the search for dislocations reveals the weakest point of
the parametrization which performs very well in other cases
(c.f. figure~\ref{rhovar}). What actually counts is not how the
parametrization works for some configurations that were specially
sought for their bad performance, but how well it performs for
configurations in thermal equilibrium occurring in a Monte Carlo
simulation. The results of a test of this performance --- presented in
section~\ref{results} --- shows that indeed there is no problem.

We also searched for the minimal action configuration in the $Q=1$
sector and did not find a configuration with an action below 
the continuum action. This is not astonishing as there are no
one--instanton solutions on a torus. 

\section{Topological Susceptibility}
\label{topological}

If the topological susceptibility is a well defined physical quantity
that is renormalization group invariant, then one expects that it
scales like a $\rm (mass)^2$ in the continuum limit. One additionally
measures a second quantity, e.g. the correlation length $\xi$ and
builds the dimensionless product $\chi_t \, \xi^2$ which should go to
a constant in the limit $\xi \rightarrow \infty$.  Earlier
Monte Carlo calculations do not show convincingly whether this is the
case.  Furthermore, perturbative considerations indicate that in
the O(3) model there might be a problem with the topological
susceptibility in the continuum limit.

One may calculate the contribution of instantons in the continuum
using a semiclassical expansion.  The probability density to find an
instanton with topological charge $Q = 1$ and size $\rho \ll 1/
\Lambda$ is \cite{JEVICKI,SCHWAB}:
\begin{equation}
P_1 \sim  \frac{d\rho}{\rho},
\label{scaleint}
\end{equation}
where $\Lambda$ is the scale parameter of the model.  Using the
renormalization group one can show, that eq.~(\ref{scaleint}) is exact
as $\rho \rightarrow 0$ \cite{LUSCHER1} assuming the small instantons
form a dilute gas. The infrared divergence at $\rho = 0$ in
eq.~(\ref{scaleint}) indicates, however, that this assumption is not
true: small instantons are not suppressed, but contribute strongly to
the susceptibility.  Such a dominance of small instantons is also
indicated by numerical studies, trying to determine the instanton size
distribution with different methods
\cite{MICHAEL,FARCHIONI1}.

On the lattice, however, not the whole range of the instanton radius
is probed. The lattice cuts the contribution of small instantons
because there is a smallest possible radius before the instanton falls
through the lattice (c.f. figure~\ref{rhovar}). Hence the measured
topological susceptibility is always finite.  It is not excluded,
however, that it raises boundlessly with increasing correlation length
as the perturbative considerations suggest.

\subsection{Numerical Results}
\label{results}

Using the perfect lattice action and the perfect charge we performed
extensive Monte Carlo simulations at correlation lengths in the range
$\xi \in (2 - 60)$. 
In order to avoid finite size effects we kept the ratio $L/\xi\approx 6$
constant. The correlation length was obtained from the long distance
fall off of the zero momentum correlation function. 
 
We determined the topological charge using both the geometric
definition and the definition of the FP charge given in
section~\ref{tcharge}. For the measurement of the FP charge we used the
geometric definition of the charge on a finer lattice of the multigrid
with the Monte Carlo generated lattice as coarsest level.  
To determine the configuration $\Sv(\Rv)$ on the fine lattice one can 
either minimize the FP equation (which is very time consuming) or use 
the parametrization of the dependence on $\Rv$ given in
section~\ref{parfield}. We denote the corresponding charges $Q_{\rm
coarse}$ for the geometric charge, $Q_{\rm par.\,1.\,level}$
for the charge measured on the first finer level using the
parametrization of the fine field etc.
  
For two $\beta$ values we compared the results of using the
parametrization on a finer level and of minimizing on a multigrid. 
As is shown in table~\ref{test}, the
results were found to be consistent within the statistical errors. 
This confirms, that the parametrization
performs well for configurations occurring in a Monte Carlo simulation
(c.f. the discussion at the end of section~\ref{classres}).
In order to test if the result of going to a lower level is already stable,
we also calculated the charge for a few $\beta$ values 
on the second finer lattice. As also can be seen in table~\ref{test}, the
values on the first and the second finer level were found to be
consistent within the statistical errors.
Thus it is sufficient to calculate the fine field only
on the first finer level. This is not 
astonishing as the maximal angle between two neighbouring spins halves
as one goes one step down to a finer level. So even on the first finer
level the maximal possible angle between two neighbouring spins is
$90^{\circ}$ and there is practically no ambiguity left for the
topological charge. 

\begin{table}\centering
\begin{tabular}{|*{6}{r@{}l|}}
\hline
\multicolumn{2}{|c|}{$\beta$} & 
\multicolumn{2}{|c|}{$\langle Q^2_{\rm coarse} \rangle $} & 
\multicolumn{2}{|c|}{$\langle Q^2_{\rm par.\, 1.\, level} \rangle$} & 
\multicolumn{2}{|c|}{$\langle Q^2_{\rm par.\, 2.\, level} \rangle$} & 
\multicolumn{2}{|c|}{$\langle Q^2_{\rm min.\, 1.\, level} \rangle$} & 
\multicolumn{2}{|c|}{$\langle Q^2_{\rm min.\, 2.\, level} \rangle$} \\ 
\hline
0.&6   &  2.&56(5)   & 1.&96(4)    & 1.&93(4)   &   &        &   &      \\
0.&7   &  3.&39(4)   & 2.&48(3)    & 2.&45(3)   & 2.&45(3)   & 2.&45(3) \\
0.&85  &  5.&62(11)  & 4.&31(9)    & 4.&30(9)   &   &        &   &      \\
1.&0   &  6.&38(8)   & 4.&86(6)    & 4.&82(6)   &   &        &   &      \\
1.&0   &  6.&33(6)   & 4.&78(5)    &   &        & 4.&75(5)  &   &      \\ 
\hline
\end{tabular}
\caption{Results of test MC Simulations in the O(3) model indicating that it
is sufficient to measure the charge only on the first finer parametrized
level. Minimizing or going to a lower level yields the same result.}
\label{test}
\end{table}

In table~\ref{mcresults} we report the results of the simulations for
the correlation length $\xi$, the geometric charge $\langle Q_{\rm 
coarse}^2 \rangle$ and the FP charge $\langle Q^2_{\rm 1. level}
\rangle$ evaluated on the first parametrized finer level. Using the
geometric charge as well as the FP charge, we build the dimensionless
quantity $\chi^t\cdot \xi^2$ of topological
susceptibility and correlation length.

\begin{table}\centering
\begin{tabular}{|r@{.}l|r|*{5}{r@{.}l|}}
\hline
\multicolumn{2}{|c|}{$\beta$} & L & \multicolumn{2}{c|}{$\xi$} & 
\multicolumn{2}{|c|}{$\langle Q^2_{\rm coarse} \rangle $} & 
\multicolumn{2}{|c|}{$\chi^t_{\rm coarse} \cdot \xi^2$} & 
\multicolumn{2}{|c|}{$\langle Q^2_{\rm 1. level} \rangle$} & 
\multicolumn{2}{|c|}{$\chi^t_{\rm 1. level} \cdot \xi^2$} \\
\hline
0&51  &  10 &  1&6049(42) &  1&87(1)   & 0&0482(4)  & 1&329(8)  & 0&0342(3)  \\
0&6   &  14 &  2&1960(46) &  2&57(2)   & 0&0631(6)  & 1&90(2)   & 0&0467(4)  \\
0&685 &  20 &  3&012(14)  &  3&57(3)   & 0&0810(10) & 2&65(2)   & 0&0601(7)  \\
0&7   &  20 &  3&186(15)  &  3&36(3)   & 0&0852(11) & 2&52(2)   & 0&0640(8)  \\
0&85  &  40 &  6&057(17)  &  5&73(4)   & 0&1314(12) & 4&38(3)   & 0&1004(9)  \\
1&0   &  70 & 12&156(34)  &  6&36(5)   & 0&1918(19) & 4&80(4)   & 0&1448(15) \\
1&1   & 130 & 20&397(86)  & 10&04(12)  & 0&2472(36) & 7&69(9)   & 0&1893(27) \\
1&2   & 180 & 34&44(30)   &  8&23(14)  & 0&3013(73) & 6&11(10)  & 0&2237(53) \\
1&3   & 340 & 58&06(37)   & 12&01(29)  & 0&3502(96) & 8&94(22)  & 0&2607(72) \\
\hline
\end{tabular}
\caption{Results of MC Simulations with the FP action.}
\label{mcresults}
\end{table}

\begin{figure}[htb]
\begin{center}
\leavevmode
\epsfxsize=110mm
\epsfbox{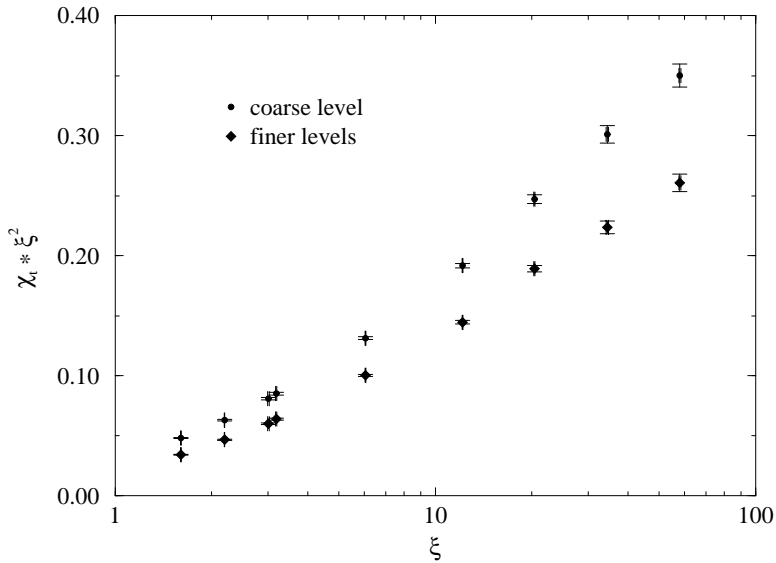}
\vskip -5mm
\caption{Results of Monte Carlo measurements of the topological
susceptibility at different correlation lengths.}
\label{noscaling}
\end{center}
\end{figure}

Figure~\ref{noscaling} shows the results for the topological
susceptibility. Clearly, no scaling is seen even at correlation
lengths as large as 60. Both curves, the one for the geometric charge
and the one for the FP charge, are rising and no flattening occurs at
the largest correlation lengths.  There is a significant difference
between the topological susceptibility built with the geometric charge
and the one with the FP charge. 
The value of the geometric charge lies several standard deviations
above the value of the FP charge. Furthermore, the difference is slowly
growing with increasing correlation length.

\section{Conclusion and Outlook}
\label{discussion}

The  definition of the FP topological charge is based on the FP field
operator.  The FP field can be evaluated to any precision desired by
solving the iterated FP equation on a $k$ level multigrid.  As was
demonstrated by fig.~\ref{rhovar} and can be shown analytically for
sufficient large multigrids the FP charge together with the FP action
has no lattice defects whatsoever.  

For use in MC simulations however a parametrization of the action and
the charge are needed.  Instead of parametrizing the charge directly,
we have parametrized the solution of the FP equation which is to be
iterated to obtain the FP field.  The accuracy of the parametrizations
have been rechecked in MC simulations.

The partition function of the lattice $\sigma$--model is --- as
indicated by a semiclassical approximation \cite{JEVICKI} ---
dominated by small sized topological excitations. These unphysical
fluctuations are the cause for the divergence which is seen in
fig.~\ref{noscaling}.

Although we studied the $\sigma$--model in this paper, the methods
derived are quite general and can be applied to other asymptotically
free theories such as $\mbox{CP}^{\mbox{\rm \tiny N-1}}$ models or
SU(N) gauge models.  For instance, in $\mbox{CP}^{\mbox{\rm \tiny
N-1}}$ models a lowest order semiclassical approximation estimates for
instanton contributions \cite{JEVICKI,SCHWAB}
\begin{equation}
I \sim \int \, d\rho \, \rho^{\mbox{\tiny N}-3}.
\end{equation}
If $\mbox{N} \geq 4$, the contribution of short distance topological
excitations is small relative to that of the physical ones and we
expect to see a scaling of the topological susceptibility according to
the perturbative RG.  In fact in the $\mbox{CP}^{3}$ model, which is
studied by one of us
\cite{BURKHALTER}, the topological susceptibility already exhibits the
expected scaling behaviour.  

\noindent{\large \bf Acknowledgments }

\noindent We wish to thank T.~DeGrand, A.~Hasenfratz, A.~Papa and
U.~Wiese for useful discussions.


\eject

\end{document}